\documentclass[12pt,a4paper,final]{revtex4}

\usepackage{sidecap}
\usepackage{ulem}
\usepackage{epsfig}
\usepackage{amsmath,amssymb,amsthm}

\usepackage{bm}
\usepackage{color}
\usepackage{xcolor}

\usepackage{enumerate}

\usepackage{float}
\usepackage{tabularx}

\usepackage{hyperref} 

\DeclareMathAlphabet{\mathpzc}{OT1}{pzc}{m}{it}

\usepackage{graphicx}

\begin{document}

\renewcommand{\textfraction}{0.00}


\newcommand{\vAi}{{\cal A}_{i_1\cdots i_n}}
\newcommand{\vAim}{{\cal A}_{i_1\cdots i_{n-1}}}
\newcommand{\vAbi}{\bar{\cal A}^{i_1\cdots i_n}}
\newcommand{\vAbim}{\bar{\cal A}^{i_1\cdots i_{n-1}}}
\newcommand{\htS}{\hat{S}}
\newcommand{\htR}{\hat{R}}
\newcommand{\htB}{\hat{B}}
\newcommand{\htD}{\hat{D}}
\newcommand{\htV}{\hat{V}}
\newcommand{\cT}{{\cal T}}
\newcommand{\cM}{{\cal M}}
\newcommand{\cMs}{{\cal M}^*}
\newcommand{\vk}{\vec{\mathbf{k}}}
\newcommand{\bk}{\bm{k}}
\newcommand{\kt}{\bm{k}_\perp}
\newcommand{\kp}{k_\perp}
\newcommand{\km}{k_\mathrm{max}}
\newcommand{\vl}{\vec{\mathbf{l}}}
\newcommand{\bl}{\bm{l}}
\newcommand{\bK}{\bm{K}}
\newcommand{\bb}{\bm{b}}
\newcommand{\qm}{q_\mathrm{max}}
\newcommand{\vp}{\vec{\mathbf{p}}}
\newcommand{\bp}{\bm{p}}
\newcommand{\vq}{\vec{\mathbf{q}}}
\newcommand{\bq}{\bm{q}}
\newcommand{\qt}{\bm{q}_\perp}
\newcommand{\qp}{q_\perp}
\newcommand{\bQ}{\bm{Q}}
\newcommand{\vx}{\vec{\mathbf{x}}}
\newcommand{\bx}{\bm{x}}
\newcommand{\tr}{{{\rm Tr\,}}}
\newcommand{\bc}{\textcolor{blue}}

\newcommand{\beq}{\begin{equation}}
\newcommand{\eeq}[1]{\label{#1} \end{equation}}
\newcommand{\ee}{\end{equation}}
\newcommand{\bea}{\begin{eqnarray}}
\newcommand{\eea}{\end{eqnarray}}
\newcommand{\beqar}{\begin{eqnarray}}
\newcommand{\eeqar}[1]{\label{#1}\end{eqnarray}}

\newcommand{\half}{{\textstyle\frac{1}{2}}}
\newcommand{\ben}{\begin{enumerate}}
\newcommand{\een}{\end{enumerate}}
\newcommand{\bit}{\begin{itemize}}
\newcommand{\eit}{\end{itemize}}
\newcommand{\ec}{\end{center}}
\newcommand{\bra}[1]{\langle {#1}|}
\newcommand{\ket}[1]{|{#1}\rangle}
\newcommand{\norm}[2]{\langle{#1}|{#2}\rangle}
\newcommand{\brac}[3]{\langle{#1}|{#2}|{#3}\rangle}
\newcommand{\hilb}{{\cal H}}
\newcommand{\pleft}{\stackrel{\leftarrow}{\partial}}
\newcommand{\pright}{\stackrel{\rightarrow}{\partial}}

\newcommand{\squeezeup}{\vspace{-2.5mm}}

\newcommand{\RomanNumeralCaps}[1]
    {\MakeUppercase{\romannumeral #1}}


\title{Understanding mass hierarchy in collisional energy loss through heavy flavor data}

\date{\today}

\author{Bojana Ilic$^1$ and Magdalena Djordjevic$^1$}

\affiliation{$^1$Institute of Physics Belgrade, University of Belgrade, Belgrade, Serbia}

\begin{abstract}
While experimental observations, such as the mass hierarchy effect, are attributed and analyzed within radiative models, their interpretation crucially depends on collisional energy loss contribution, which is often neglected in such analyses. To our knowledge, neither a (direct) simple relation between collisional energy loss and heavy quark mass is established, nor an observable that quantifies this effect. On the other hand, the upcoming high-luminosity measurements at RHIC and LHC will generate heavy flavor data with unprecedented precision, providing an opportunity to utilize high-$p_\perp$ heavy flavor data to analyze the interaction mechanisms in the quark-gluon plasma. To this end, we employ a recently developed DREENA framework based on our dynamical energy loss formalism to study the mass hierarchy in heavy flavor suppression. We present i) Analytical derivation of a direct relation between collisional suppression/energy loss and heavy quark mass. ii) A novel observable sensitive only to the collisional energy loss mechanism to be tested by future high-precision experiments. iii) Analytical and numerical extraction of the mass hierarchy in collisional energy losses through this observable.
\end{abstract}

\maketitle
\section{Introduction}
For over two decades, ultra-relativistic heavy-ion collisions at the RHIC and the LHC have been aiming to create and understand the features of the new form of matter $-$ the quark-gluon plasma~\citep{QGP0,QGP2,QGP3,QGP4} (QGP). The rare high-$p_\perp$ partons, while traversing the medium, interact with the QGP bulk and lose energy, which results in jet quenching~\citep{Bj}. The jet quenching is recognized as one of the most important probes of the QGP medium, with a crucial role in the QGP discovery~\citep{discoveryQGP}. Today, the joint theoretical and experimental effort is necessary for providing unbiased insight into the nature of parton-medium interactions and, consequently, the QGP features. Within this, an important goal presents a search for adequate effect and observable for characterizing the appropriate energy loss mechanisms.

Due to the prevailing opinion that gluon's bremsstrahlung is the dominant mechanism of high-$p_\perp$ parton energy loss~\citep{LCPI,BDMPS,ASW,QW,GLV,AMY,HT,HT1}, many experimental observations~\citep{MOexp} are attributed and analyzed within radiative models. On the other hand, in the intermediate-$p_\perp$ regime ($p_\perp \lesssim 10$ GeV) it is considered that, due to the dead-cone effect~\citep{DCE}, elastic interactions prevail for heavy flavor~\citep{Svet,MooreT,Rapp,Pol,Hirano,Molnar0}. Moreover, in~\citep{PLF,Coll,eff,collBitno1,Mustafa,collBitno2}, it was shown that for the charm and bottom quarks, the collisional energy loss is comparable to, or even larger, than radiative energy loss.

A major research goal has been to understand the mass hierarchy in parton energy loss. In the case of inelastic scattering, it is known as the dead-cone effect~\citep{DCE}, which reflects in gluon radiation suppression of the heavy (i.e., bottom) quarks compared to the light quarks at small angles. While the dead-cone effect is extensively studied, both analytically and numerically, within different radiative energy loss models~\citep{DjG,MradDB,ASW,DCE3,rad0}, the mass hierarchy in collisional energy loss is not yet addressed.
Within this, a specific challenge presents: {\it i)} A search for an observable, which can disentangle collisional from radiative energy loss; {\it ii)} Analytical derivation of a direct relation between collisional suppression/energy loss and heavy quark mass. With this goal in mind, we here propose, through analytical and numerical analysis within our DREENA-C framework~\citep{DREENAC}, a novel observable sensitive to the mass hierarchy in collisional energy loss of high-p$_\perp$ particles. Further, we demonstrate that our estimate for mass hierarchy in collisional energy loss is in qualitatively and quantitatively good agreement with the existing (scarce) experimental data. While current error bars at the RHIC and the LHC are large, we expect that the upcoming high-precision measurements will be able to directly extract mass dependence in collisional energy loss from the data.

\section{Computational framework}

For generating predictions, we employ our fully optimized DREENA-C~\citep{DREENAC} (Dynamical Radiative and Elastic ENergy loss Approach, where C stands for constant/average temperature profile) framework. We opt for DREENA-C instead of hydrodynamically-wise more upgraded versions DREENA-B~\citep{DREENAB} (Bjorken~\citep{BjorkenH} medium expansion) and -A (Adaptive, i.e., arbitrary temperature profile)~\citep{DREENAA}, to avoid unnecessary complications stemming from subtleties of medium evolution and consequently allow analytical derivations. This can be done without significant loss of accuracy, as we previously demonstrated that energy loss-sensitive observable $R_{AA}$ (considered in this study) is practically unaffected by the medium evolution model (for details, see Section III).

The framework consists of: i) Initial quark momentum distribution~\citep{ID}; ii) Energy loss probability rooted in the dynamical energy loss formalism~\citep{Rad,Coll,DELF}, which comprises multi-gluon~\citep{MGF} and path-length fluctuations~\citep{PLF}. The path-length fluctuations are calculated following the procedure provided in~\citep{Dainese,DREENAC}. The average temperature for each centrality bin is evaluated following the procedure outlined in Refs.~\citep{DREENAC,nonC}. It is worth noting that our predictions are valid for $p_\perp \gtrsim 7$ GeV.

The medium modified distribution of high-$p_\perp$ heavy flavor particles is obtained by utilizing the generic pQCD convolution formula~\citep{DELF,PLF}:
\begin{align}~\label{conv}
\frac{E_f d^3 \sigma_f}{dp^3_f}=\frac{E_i d^3 \sigma_i (Q)}{dp^3_i} \otimes P(E_i \rightarrow E_f),
\end{align}
where subscripts $i$ and $f$ stand for {\it initial} and {\it final} quantities, while $\frac{E_i d^3 \sigma_i}{dp^3_i}$ denotes initial heavy quark distribution computed according to~\citep{ID}. $P(E_i \rightarrow E_f)$ represents energy loss probability, based on dynamical energy loss formalism (see below). In distinction to Refs.~\citep{DELF,PLF}, Eq.~\eqref{conv} does not include the fragmentation function ($D(Q \rightarrow H_Q)$) for both charm and beauty (into $D$ and $B$ mesons), as D/B suppression presents clear (genuine) charm/bottom probe's suppression~\citep{HeavyBareC,HeavyBareC1, HeavyBare}. Also, when providing predictions for $b$ quark observables, we compare with indirect, non-prompt $J/ \psi$ total $R_{AA}$, due to its broader availability. This is legit, since due to the interplay of collisional and radiative energy losses, $B$ meson suppression is almost independent on $p_\perp$~\citep{rad0} (in the relevant region), so the fragmentation/decay functions will not play a notable role for different types of $b$ quark observables.

DREENA-C~\citep{DREENAC} is based on our dynamical energy loss formalism~\citep{Rad,Coll,DELF}, which includes several realistic features: {\it i)} QCD medium of finite size and finite temperature. {\it ii)} The medium consists of dynamical (i.e., moving) as opposed to static partons, which allows the longitudinal momentum exchange with the medium constituents; this is contrary to the medium models with widely applied vacuum-like propagators and/or static approximation~\citep{BDMPS,ASW,QW,GLV,AMY,HT}. {\it iii)} Calculations based on generalized Hard-Thermal-Loop approach~\citep{Kapusta}, where infrared divergences are naturally regulated ~\citep{Rad,Coll}. {\it iv)} Consistent inclusion of both collisional~\citep{Coll} and radiative~\citep{Rad} energy loss in the same theoretical framework. {\it v)} Incorporation of finite parton’s mass~\citep{DjG}, broadening the formalism applicability from light toward heavy flavor; {\it vi)} Generalization toward running coupling~\citep{DELF}, finite magnetic mass~\citep{Magm} and beyond soft-gluon approximation~\citep{bsga}. In Ref.~\citep{eff} we demonstrated that all these features are necessary for reliable suppression predictions.

High-$p_\perp$ $R_{AA}$ predictions, generated with DREENA-C for a large amount of experimental data at the RHIC and the LHC, show a good agreement~\citep{DELF,nonC,DREENAC} with the existing data; explain puzzling observations (such as heavy-flavor puzzle~\citep{HeavyBareC,HFP1}) and provide nonintuitive predictions for the upcoming experiments~\citep{rad0,51,masstomography} (some of which were subsequently confirmed by data~\citep{confirm1}). Consequently, our framework/formalism can adequately describe high-$p_\perp$ parton-medium interactions, and it presents a suitable framework for study conveyed in this paper. Furthermore, we use no fitting parameters in generating predictions, i.e., all the parameters correspond to standard literature values, stated in~\citep{DREENAC}. To name the most relevant ones for this study: the charm (bottom) mass is $M_c =1.2$ GeV ($M_b =4.75$ GeV). Different non-perturbative calculations~\citep{xb1,xb2} indicate chromomagnetic to chromoelectric mass ratio in the range $0.4 < \frac{\mu_B}{\mu_E} < 0.6$ for RHIC and LHC. We here opt for a $\frac{\mu_B}{\mu_E} =0.4$, while we checked that introducing a larger value has a negligible impact on our predictions.

\section{Results and Discussion}

\begin{figure}
\includegraphics[width=\linewidth]{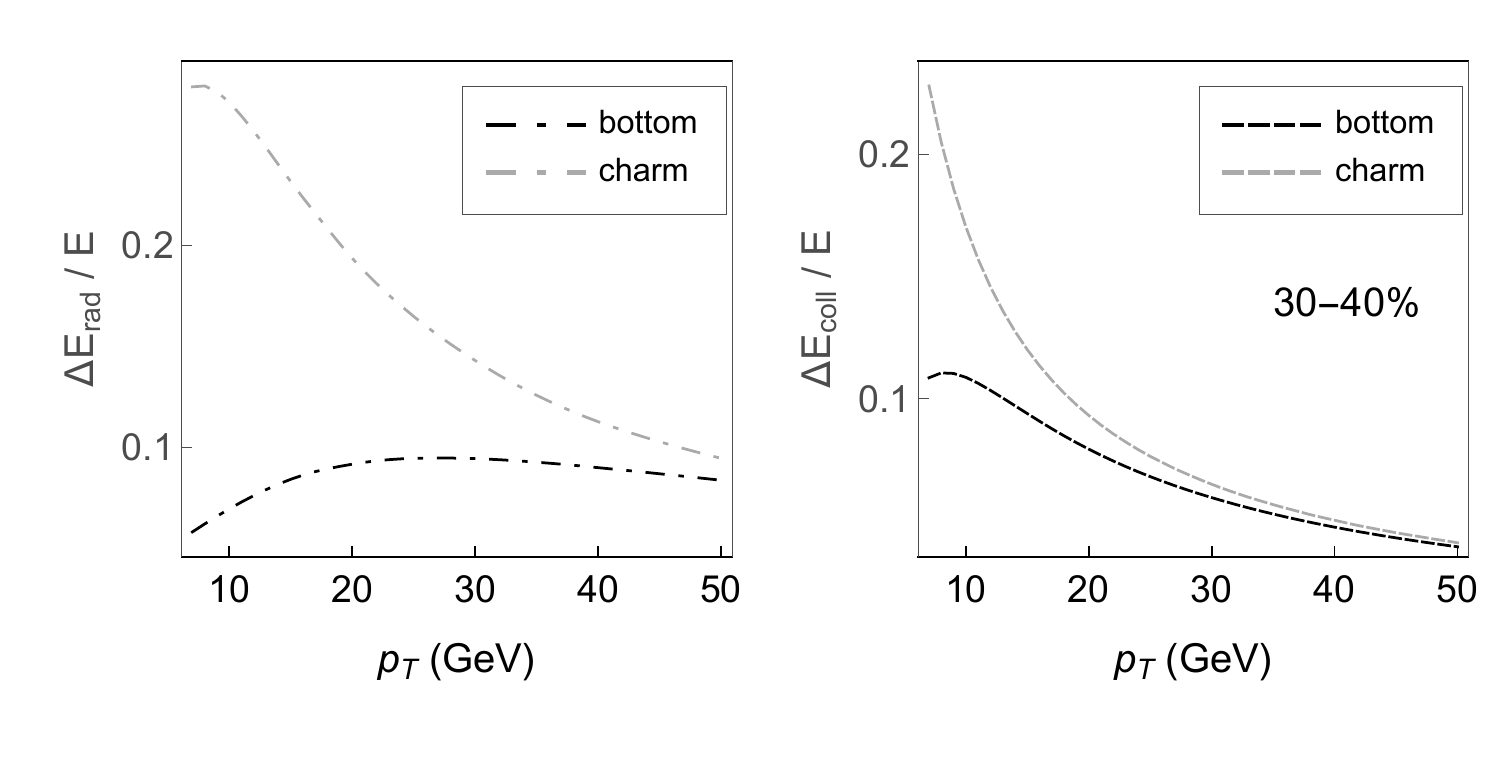}
\caption{The mass hierarchy in fractional energy losses.  Heavy quark fractional energy loss as a function of $p_\perp$, when only radiative (the left plot) and only collisional (the right plot) energy loss mechanism is assumed.  The selected centrality bin is $30-40\%$.  Black (gray) curves correspond to bottom (charm) quark.}
\label{fig:sl0}
\end{figure}

In this section, we start with comparing patterns in energy loss of charm and bottom quarks within the DREENA-C framework~\citep{DREENAC}. From the left plot of Fig.~\ref{fig:sl0}, we reproduce the dead-cone effect, i.e., mass hierarchy in radiative energy loss. Namely, we see that the bottom quark, due to larger mass, loses significantly less energy compared to the charm quark~\citep{MradDB,rad0}. This is especially pronounced at lower $p_\perp \approx 10$ GeV, where $M_b$ is comparable to the transverse momentum (i.e., energy, since we focus on midrapidity). This difference in $\frac{\Delta E_{rad}}{E}$ between bottom and charm decreases with increasing transverse momentum, which can be attributed to the fact that for both flavors, the mass becomes negligible compared to their $p_\perp$. Thus, already at $p_\perp \approx 50$ GeV, these two curves approach each other in accordance with~\citep{masstomography}. Though we primarily show centrality bin $30-40\%$, the results are the same regardless of the selected centrality range and therefore omitted.

For collisional energy loss, shown in the right plot of Fig.~\ref{fig:sl0}, we observe a qualitatively similar tendency. That is, we obtain clear mass hierarchy in $\frac{\Delta E_{coll}}{E}$ (see also e.g.~\citep{Coll,PollMarlene,KcollH,collAyala,Bleicher}), with heavier quark encountering notably smaller collisional energy loss at $p_\perp \approx 10$. Compared to the fractional radiative energy loss, this effect is less pronounced (and disappears faster with increasing $p_\perp$), but it is an important observation.

\begin{figure}
\includegraphics[width=0.6\linewidth]{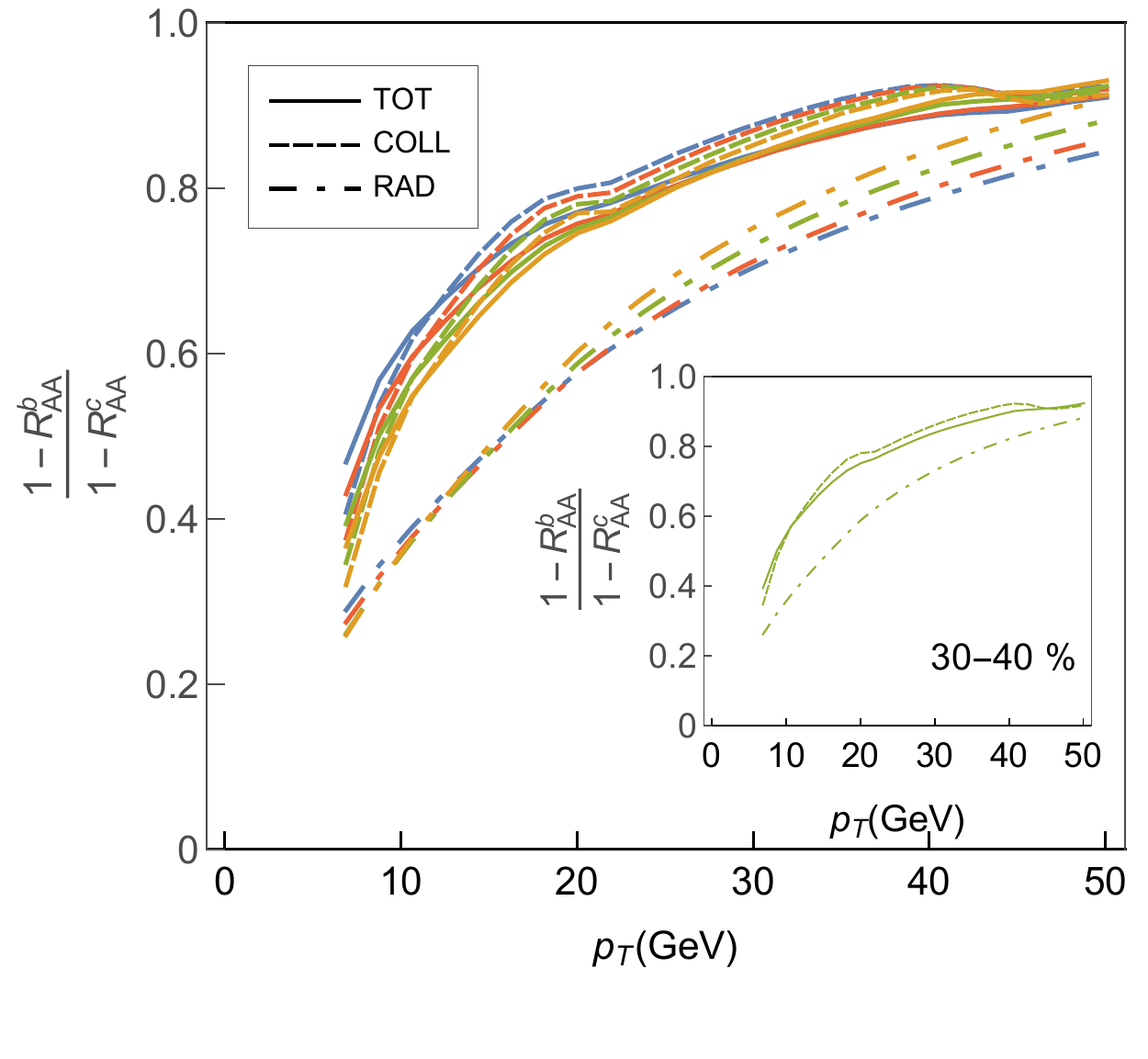}
\caption{Comparison of $1 - R_{AA}$ bottom to charm ratios for {\it total, collisional} and {\it radiative} suppressions, generated with DREENA-C~\citep{DREENAC} framework. For clarity, $30-40 \%$ centrality is presented in the inset. Full, dashed and dot-dashed curve denotes {\it total, collisional} and {\it radiative} case, respectively, as indicated in the legend. The blue, red, green and orange curves correspond to $10-20 \%$, $20-30 \%$, $30-40 \%$, and $40-50 \%$ centrality bin, respectively.}
  \label{fig:sl2}
\end{figure}

To quantify this effect on the experimental observables, we recall that $R_{AA}$ is well suited for our study, as it is susceptible to energy loss~\citep{eff} while being practically insensitive to the details of medium evolution (contrary to, e.g. elliptic flow)~\citep{DREENAC,DREENAB, PathD, RAAobs1,RAAobs2}. Therefore, it is reasonable to assume that the adequate observable should be a function of $R_{AA}$. In particular, we will further analyze $1-R_{AA}$, as this observable carries more direct information on the parton energy loss than commonly used $R_{AA}$~\citep{PathD}. To this end, in Fig.~\ref{fig:sl2}, we compare $1-R_{AA}$ bottom to charm ratios, when {\it total} (radiative and collisional) energy losses are accounted for. In the same plot, we also separately consider the $1-R_{AA}$ ratio, resulting from only {\it collisional} and only {\it radiative} energy loss scenarios. The predictions for all considered centrality ranges are displayed.

Fig.~\ref{fig:sl2} contains a large number of curves, which makes it overcrowded and may obscure some important observations. Therefore, an inset corresponding only to $30-40 \%$ centrality is added for transparency. Note that the choice of centrality does not influence the conclusion, as we checked that we observe the same for other centralities.
The inset provides a somewhat unexpected insight. That is, we observe that $1- R^{coll}_{AA}$ bottom to charm ratio is practically overlapping with $1- R^{tot}_{AA}$ ratio. Furthermore, {\it total} (and {\it collisional} likewise) ratio is notably larger than $1- R^{rad}_{AA}$ bottom to charm ratio. As expected, the suppression curve corresponding to the {\it total} energy loss is in between the {\it collisional} and {\it radiative} energy loss scenarios. It is, however, surprising that the {\it total} energy loss curve closely follows the {\it collisional} energy loss curve, i.e., that the {\it radiative} contribution is negligible. Consequently, this figure demonstrates that the $1- R_{AA}$ ratio for heavy flavor is dominantly driven by mass hierarchy from {\it collisional} contribution.

To analytically analyze what is reflected by the ratio in Fig.~\ref{fig:sl2}, we start from the definition of high-$p_\perp$ particle suppression, assuming only collisional interactions within the QGP. To obtain the final particle spectrum ($d \sigma^f/ d p^2_{\perp}$) at midrapidity, the standard procedure~\citep{BDMS} is a convolution of the initial parton momentum distribution ($d \sigma^i (p_{\perp}+\epsilon) /{d p^2_{\perp}}$) with the energy loss probability ($D(\epsilon)$) in the final stage~\citep{Mustafa}. The assumption that energy loss of a high-$p_\perp$ heavy flavor is small (i.e., $\epsilon \ll p_\perp$) allows Taylor expansion in:
\begin{align}~\label{InitDt}
\frac{d \sigma^f}{d p^2_{\perp}} &{} = \int{d \epsilon D(\epsilon) \frac{d \sigma^i (p_{\perp}+\epsilon)}{d p^2_{\perp}}} \nonumber \\
& = \int{d \epsilon D(\epsilon) \frac{d \sigma^i (p_{\perp})}{d p^2_{\perp}}} +\int{d \epsilon D(\epsilon) \ \frac{\epsilon}{1!} \ \frac{d}{d p_\perp} \Big( \frac{d \sigma^i (p_{\perp})}{d p^2_{\perp}}} \Big) + ...\nonumber \\
& \simeq  \frac{d \sigma^i}{d p^2_{\perp}} + \Delta E_{coll} \frac{d}{d p_\perp} \Big( \frac{d \sigma^i}{d p^2_{\perp}}\Big).
\end{align}
Here, we use that the probability $\int{d \epsilon D(\epsilon)}=1$, as well as the fact that the total collisional energy loss of parton in the medium is given by $\Delta E_{coll}=\int{d \epsilon D(\epsilon)} \epsilon$.

\begin{figure}
\includegraphics[width=0.9\linewidth]{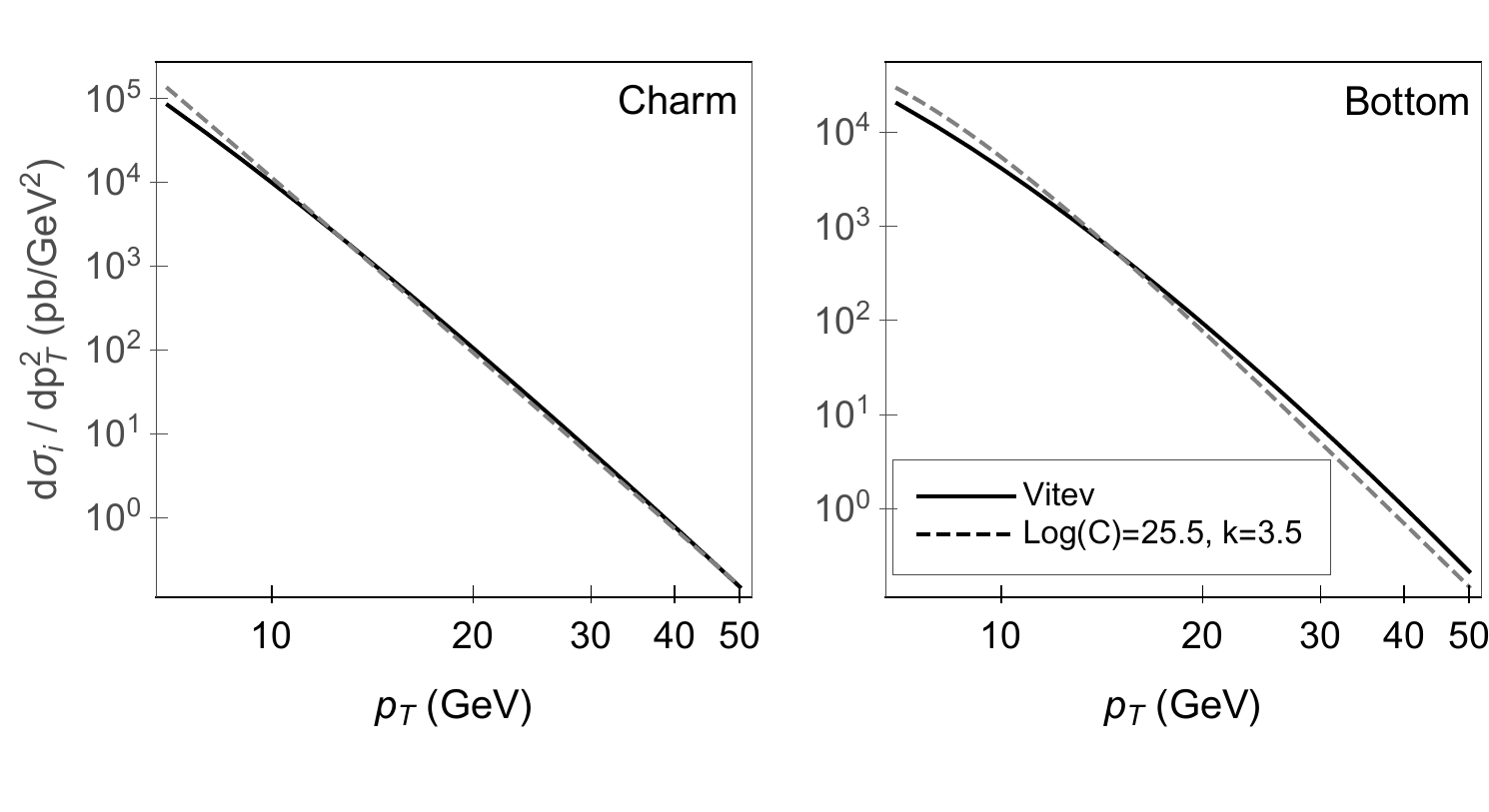}
\caption{Mass parameterization of transverse momentum distributions for charm and bottom.  Charm and bottom initial distributions are presented on the left and the right plot, respectively, as a function of $p_\perp$. On each plot, full black curve corresponds to the distribution, computed at next to leading order according to~\citep{ID}, while gray dashed curve represents our mass-dependent fitted distribution, based on Eq.~\eqref{MTe}.}
  \label{fig:sl1}
\end{figure}

Furthermore, we assume that initial $p_\perp$ distribution of a heavy parton can be parameterized as following~\citep{MooreT}:
\begin{align}~\label{MTe}
\frac{d \sigma^i}{d p^2_{\perp}} = \frac{C}{(p^2_\perp+M^2)^k},
\end{align}
where constants $C$ and $k$ should be the same for the charm and bottom quark. Indeed, for the used initial distributions~\citep{ID}, we explicitly verified this equality in Fig.~\ref{fig:sl1}, where our numerical analysis yields the consistent values for both heavy partons ($\ln ( C) \simeq 25.5$, $k\simeq 3.5$).

After taking derivative of Eq.~\eqref{MTe} with respect to $p_\perp$, Eq.~\eqref{InitDt} straightforwardly reduces to:
\begin{align}~\label{veza}
\frac{d \sigma^f}{d p^2_{\perp}} &{}\simeq   \frac{d \sigma^i}{d p^2_{\perp}} \Big (1-2 k \  \frac{p_\perp}{E} \ \frac{\Delta E_{coll}}{E} \Big),
\end{align}
where $E=\sqrt{p^2_\perp +M^2}$. Since parton's suppression~\citep{QW} is defined as~\citep{MooreT} $R_{AA}=\frac{ d \sigma^f/dp^2_\perp}{d \sigma^i/dp^2_\perp}$, we finally obtain:
\begin{align}~\label{proporcionalnost}
1-R_{AA} &{} \simeq 2 k \  \frac{p_\perp}{E} \ \frac{\Delta E_{coll}}{E}.
\end{align}

To extract the analytic dependence of $1-R_{AA}$ on the mass of heavy quark, we need to analyze the mass dependence of collisional energy loss analytically. Since our energy loss expression~\citep{Coll} is highly nontrivial and not analytically tractable, we opt for a more straightforward Thoma-Gyulassy~\citep{TG} result, which is moreover in a (reasonably) good agreement with our result~\citep{Coll} (in the $p_\perp$ range of concern). After algebraic manipulation, we obtain that the proportionality between fractional collisional energy loss and parton's mass is represented by (see Eq.~\eqref{dE3} in the Appendix):
\begin{align}~\label{dE31}
\frac{\Delta E_{coll}}{E} \sim \frac{1}{p_\perp} \Big( 1-\frac{M^2}{p^2_\perp} \ln{\frac{2 p_\perp}{M}} \Big).
\end{align}
Along the same lines, from Eq.~\eqref{proporcionalnost}, we obtain that the mass dependence of $1-R_{AA}$ is represented by (see Eq.~\eqref{R1} in the Appendix):
\begin{align}~\label{R11}
1-R_{AA} &{} \sim \frac{2 k}{p_\perp } \Big[ 1-\frac{M^2}{p^2_\perp} \Big( \ln{2} +\frac{1}{2} \Big)+ \Big( \frac{M}{p_\perp} \Big)^{\frac{M}{p_\perp}+1} -\frac{M}{p_\perp} \Big].
\end{align}
Surprisingly, further numerical consideration revels that the dominant mass term acquires the form:
\begin{align}~\label{RAAprop0}
1-R_{AA} &{} \sim    \frac{2 k}{p_\perp}  \Big( 1-\frac{M}{p_\perp} \Big).
\end{align}
We further form bottom to charm $1-R_{AA}$ ratio, so that a common factor $\frac{2 k}{p_\perp}$ is canceled, leading to:
\begin{align}~\label{RAAprop}
\frac{1-R^b_{AA}}{1-R^c_{AA}}  &{} \simeq \frac{ 1-\frac{M_b}{p_\perp}}{ 1-\frac{M_c}{p_\perp}}.
\end{align}
Thus, we obtain that the $1-R_{AA}$ ratio for heavy flavor is surprisingly simple, depends only on the mass and momentum of heavy quarks, and is independent of the considered centrality.

To test the centrality independence, we go back to Fig.~\ref{fig:sl2}. We see that {\it total} and {\it collisional} $1-R_{AA}$ bottom to charm ratios are nearly indistinguishable regardless of the centrality bin, i.e., as predicted, do not depend on the collision centrality.

\begin{figure}
\includegraphics[width=0.6\linewidth]{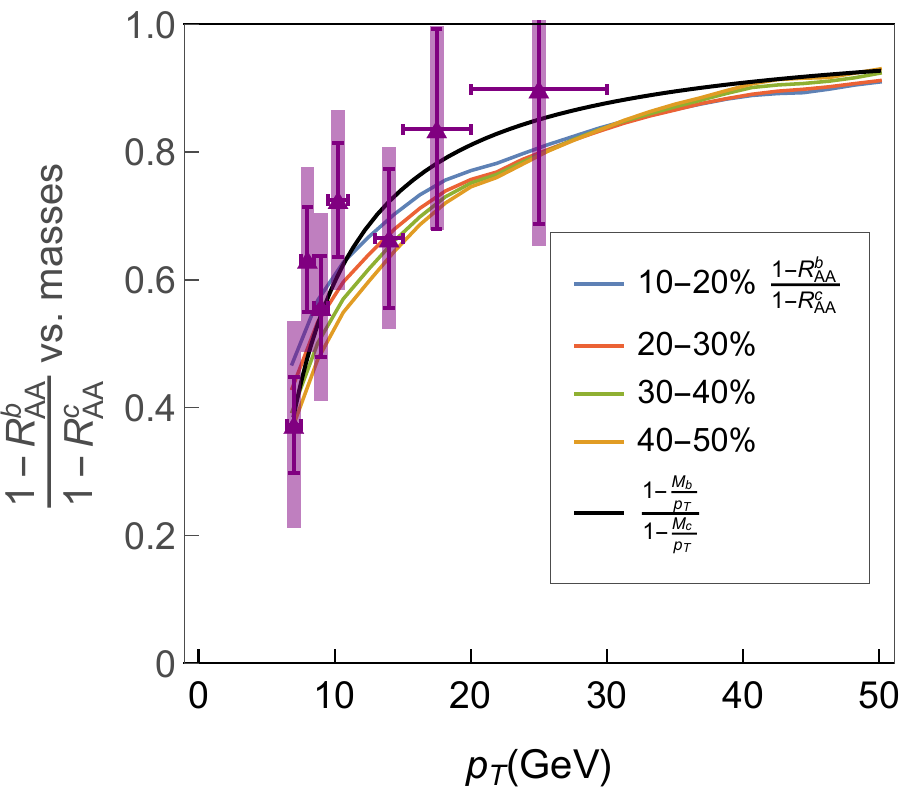}
\caption{ Quantifying the mass hierarchy in collisional energy loss. $1 - R_{AA}$  bottom to charm ratios for {\it total} suppression at $10-20 \%$ (blue curve), $20-30 \%$ (red curve), $30-40 \%$ (green curve) and $40-50 \%$ (orange curve) as a function of $p_\perp$. The predictions are generated within our full-fledged DREENA-C~\citep{DREENAC} suppression numerical procedure
and compared with the $1-R_{AA}$ ratio of non-prompt $J/\psi$ at $30-100 \%$ CMS~\citep{CMS_jpsi} and average $D^0,D^+,D^{*+}$ at $30-50 \%$ ALICE~\citep{ALICE_D} 5.02 TeV Pb+Pb collisions (purple triangles). The black curve corresponds to the extracted mass dependence (see Eq.~\eqref{RAAprop}).}
\label{fig:sl4}
\end{figure}

Finally, in Fig.~\ref{fig:sl4}, DREENA-C~\citep{DREENAC} predictions of {\it total} $1-R_{AA}$ bottom to charm ratios are compared with our analytical estimate, presented on the right-hand-side of Eq.~\eqref{RAAprop}. From this figure, we observe a good agreement between our predictions and $\frac{1-M_b/p_\perp}{1-M_c/p_\perp}$ for all considered centralities. This implies the validity of the analysis presented here. A small disagreement could be attributed to the fact that our estimate originates only from collisional energy loss/suppression expression and is in agreement with Fig.~\ref{fig:sl2}.

Furthermore, Fig.~\ref{fig:sl4} also provides experimental CMS data~\citep{CMS_jpsi,ALICE_D} for $1-R_{AA}$ ratio between non-prompt $J/\psi$ on one side, and average $D$ mesons on the other side. Due to the lack of experimental data in the same centrality bins for $b$ and $c$ probes at 5.02 TeV Pb+Pb collisions, and for consistency throughout the paper, we choose overlapping bins: $30-100 \%$ for non-prompt $J/\psi$, while $30-50 \%$ for average $D$ mesons. From this figure, we observe qualitatively and quantitatively good agreement between {\it i}) the data and our predictions, supporting the validity of the DREENA-C framework used in this study, and {\it ii}) the data and our analytical mass estimate, confirming the adequacy of the proposed observable given by Eq.~\eqref{RAAprop}, and justifying the applied approximations (see the Appendix). As the error bars are quite large, our study also implies a need for higher precision data for a more satisfactory test of the proposed observable.
Furthermore, suppression measurements for both $B$ (or non-prompt $J/\psi$) and $D$ mesons in the same centrality bins are needed for extracting the mass hierarchy in collisional energy loss from the data.

\section{Conclusions and Outlook}

One of the inherent characteristics of parton's energy loss is the apparent flavor dependence. Inspired by the dead-cone effect~\citep{DCE} in radiative energy loss and experimentally observed~\citep{MOexp} mass ordering in non-prompt $J/\psi$ ($B$) and $D$ meson suppressions, we addressed the mass hierarchy in heavy flavor suppression. We found that the $1-R_{AA}$ ratio for heavy flavor reflects the mass hierarchy in the collisional energy loss, which is a nontrivial and important result.

While the dead-cone effect is extensively studied within different radiative energy loss models~\citep{DjG,MradDB,ASW,DCE3,rad0}, the mass hierarchy in collisional energy loss was not previously addressed. To our knowledge, no direct relationship between collisional energy loss and heavy quark mass is established. To this end, the analytical results provided here yielded a simple relation between collisional suppression/energy loss and heavy quark mass. Also, through joint numerical and analytical analysis within our DREENA framework~\citep{DREENAC}, we proposed a novel observable for straightforwardly extracting the mass hierarchy in collisional energy loss through heavy flavor data to be more rigorously tested by future high-precision experiments. It is based on one of the most common jet quenching observable $-$ the high-$p_\perp$ $R_{AA}$, and is independent of the collision centrality.

As an outlook, the analysis provides specific guidelines on where future experimental efforts regarding this aim should be focused. For instance, the mass hierarchy is more pronounced at lower $p_\perp$. This momentum region is experimentally accessible for both RHIC and LHC in the upcoming high-luminosity experiments, so data from both experiments can be used to test this observable. Furthermore, it is undoubtedly useful to provide $B$ meson suppression data. Finally, the suppression measurement of both $B$ (or non-prompt $J/\psi$) and $D$ mesons in the same centrality bins would be beneficial for readily extracting mass hierarchy in collisional energy loss from the data.

\appendix

\section{Approximation of collisional energy loss and ${\mathbf{1-R_{AA}}}$}
In this Section, we first simplify the analytically tractable collisional energy loss from~\citep{TG}, by assuming $M/p_\perp \ll 1$. We start from:
\begin{align}~\label{dE1}
\frac{\Delta E_{coll}}{E} \sim \frac{1}{E v^2} \Big( v+ \frac{v^2 -1}{2} \ln{\frac{1+v}{1-v}} \Big),
\end{align}
where $v=p_\perp/ \sqrt{p^2_\perp +M^2}$ denotes magnitude of the velocity of initial parton ($\vec{v}$). We Taylor expand Eq.~\eqref{dE1} for $M/p_\perp \ll 1$. Starting from $v \simeq 1-\frac{M^2}{2 p^2_\perp}$ ($ \frac{1}{(1+x)^{\frac{1}{2}}} \simeq 1-\frac{x}{2}$), first we obtain:
\begin{align}~\label{dE2}
\frac{\Delta E_{coll}}{E} \sim \frac{1}{p_\perp} \frac{1}{\sqrt{1+\frac{M^2}{p^2_\perp} }} \Big( 1+\frac{M^2}{p^2_\perp} \Big) \Big[ 1-\frac{M^2}{2 p^2_\perp} -\frac{M^2}{2 p^2_\perp \big( 1+\frac{M^2}{p^2_\perp} \big)} \ln{\Big( \frac{4 p^2_\perp}{M^2}-1 \Big)} \Big].
\end{align}
To further simplify the above expression, we frequently use the same Taylor expansion $(1+x)^{-n}\simeq1-n x$, as well as $4 p^2_\perp/{M^2} \gg 1$, while keeping the leading terms in $M/p_\perp$ expansion. Thus, Eq.~\eqref{dE2} reduces to:
\begin{align}~\label{dE3}
\frac{\Delta E_{coll}}{E} \sim \frac{1}{p_\perp} \Big( 1-\frac{M^2}{p^2_\perp} \ln{\frac{2 p_\perp}{M}} \Big).
\end{align}
Here we encounter nontrivial term  $x^2 \ln{x}$, where $x=\frac{M}{p_\perp}$, for which we are seeking the approximation for small $x$. We start from the similar expression $x \ln{x}$, and apply the trick of raising the expression into the exponent
\begin{align}~\label{LO}
e^{x \ln{x}} =x^x,
\end{align}
where we use the logarithm rules. Note, however, that $\lim_{x\to 0} x \ln {x}=\lim_{x\to 0} \frac{\ln {x}}{1/x} =\lim_{x\to 0} \frac{1/x}{-1/x^2}=0$, where we applied L'Hospital's rule~\citep{Lop}.
Since the exponent in Eq.~\eqref{LO} is close to zero, we may Taylor expand the left-hand side of this equation, which leads to $x \ln{(x)} \simeq x^x-1$. Likewise,
\begin{align}~\label{LO1}
x^2 \ln{(x)} \simeq x^{x+1}-x.
\end{align}
By substituting Eq.~\eqref{LO1} in Eq.~\eqref{dE3} we obtain:
\begin{align}~\label{dE4}
\frac{\Delta E_{coll}}{E} \sim \frac{1}{p_\perp} \Big[ 1-\frac{M^2}{p^2_\perp} \ln{2} + \Big( \frac{M}{p_\perp} \Big)^{\frac{M}{p_\perp}+1} -\frac{M}{p_\perp} \Big].
\end{align}

Next we substitute Eq.~\eqref{dE4} in Eq.~\eqref{proporcionalnost} resulting in:
\begin{align}~\label{R1}
1-R_{AA} &{} \sim 2 k \frac{1}{\sqrt{p^2_\perp +M^2} } \Big[ 1-\frac{M^2}{p^2_\perp} \ln{2} + \Big( \frac{M}{p_\perp} \Big)^{\frac{M}{p_\perp}+1} -\frac{M}{p_\perp} \Big] \nonumber \\
& = \frac{2 k}{p_\perp } \Big[ 1-\frac{M^2}{p^2_\perp} \Big( \ln{2} +\frac{1}{2} \Big)+ \Big( \frac{M}{p_\perp} \Big)^{\frac{M}{p_\perp}+1} -\frac{M}{p_\perp} \Big],
\end{align}
where in the second line of this equation, we again utilized $ \frac{1}{(1+x^2)^{\frac{1}{2}}} \simeq 1-\frac{x^2}{2}$.

This expression can be further simplified, since we explicitly checked that the second and third terms in Eq.~\eqref{R1}, on one side, are of an opposite sign, and that their sum is much smaller compared to the remaining terms on the other side. Finally, we obtain a simple proportionality:
\begin{align}~\label{R2}
1-R_{AA} \sim \frac{2 k}{p_\perp } \Big(1-\frac{M}{p_\perp} \Big).
\end{align}

{\em Acknowledgments:}
This work is supported by the European Research Council, grant ERC-2016-COG: 725741, and by the Ministry of Science and Technological Development of the Republic of Serbia, under project No. ON171004.


\begin{references}
\bibitem{QGP0} M. Gyulassy and L. McLerran, Nucl. Phys. A {\bf 750}, 30 (2005).
\bibitem{QGP2} E. V. Shuryak, Nucl. Phys. A {\bf 750}, 64 (2005); Rev. Mod. Phys. {\bf 89}, 035001 (2017).
\bibitem{QGP3} C. V. Johnson and P. Steinberg, Physics Today {\bf 63}, 29
(2010).
\bibitem{QGP4} B. Jacak and P. Steinberg, Phys. Today {\bf 63}, 39 (2010).
\bibitem{Bj} J. D. Bjorken, FERMILAB-PUB-82-059-THY, 287 (1982).
\bibitem{discoveryQGP} J. Adams {\it et al.} [STAR Collaboration], Phys. Rev. Lett. {\bf 91}, 072304 (2003); C. Adler {\it et al.} [STAR Collaboration], Phys. Rev. Lett. {\bf 90}, 082302 (2003).
\bibitem{LCPI} B.~G. Zakharov, JETP Lett. {\bf 70}, 176 (1999); JETP Lett. {\bf 73}, 49 (2001).
\bibitem{BDMPS} R. Baier, Y. Dokshitzer, A. Mueller, S. Peigne and D.Schiff, Nucl. Phys. B {\bf 484}, 265 (1997).
\bibitem{ASW} N. Armesto, C. A. Salgado and U. A. Wiedemann, Phys. Rev. D {\bf 69}, 114003 (2004).
\bibitem{QW} C.~A.~Salgado and U.~A.~Wiedemann,
Phys. Rev. D \textbf{68}, 014008 (2003).
\bibitem{GLV} M. Gyulassy, P. Levai and I. Vitev, Nucl. Phys. B \textbf{594}, 371 (2001).
\bibitem{AMY} P. B. Arnold, G. D. Moore and L. G. Yaffe, JHEP {\bf 0206}, 030 (2002).
\bibitem{HT} X. N. Wang and X. F. Guo, Nucl. Phys. A {\bf 696}, 788 (2001).
\bibitem{HT1} A. Majumder and C. Shen, Phys. Rev. Lett. {\bf 109}, 202301 (2012).
\bibitem{MOexp} A.~M.~Sirunyan \textit{et al.} [CMS],
Phys. Lett. B \textbf{782}, 474-496 (2018).
\bibitem{DCE} Y.~L.~Dokshitzer and D.~E.~Kharzeev,
Phys. Lett. B \textbf{519}, 199-206 (2001).
\bibitem{MooreT} G.~D.~Moore and D.~Teaney,
Phys. Rev. C \textbf{71}, 064904 (2005).
\bibitem{Svet} B. Svetitsky, Phys. Rev. D {\bf 37}, 2484-2491 (1988).
\bibitem{Pol} P. B. Gossiaux and J. Aichelin, J. Phys. G {\bf 36}, 064028 (2009).
\bibitem{Rapp} H. van Hees, R. Rapp, Phys. Rev. C {\bf 71}, 034907 (2005); H. van Hees, V. Greco, R. Rapp, Phys. Rev. C {\bf 73}, 034913 (2006).
\bibitem{Hirano} Y. Akamatsu, T. Hatsuda, T. Hirano, Phys. Rev. C {\bf 79}, 054907 (2009).
\bibitem{Molnar0} D.~Molnar,
Eur. Phys. J. C \textbf{49}, 181-186 (2007).
\bibitem{PLF} S. Wicks, W. Horowitz, M. Djordjevic and M. Gyulassy, Nucl. Phys. A {\bf 784}, 426 (2007).
\bibitem{Coll} M.~Djordjevic,
Phys. Rev. C \textbf{74}, 064907 (2006).
\bibitem{eff} B. Blagojevic and M. Djordjevic, J. Phys. G {\bf 42}, 075105
(2015).
\bibitem{collBitno1} M. H. Thoma, Phys. Lett. B {\bf 273}, 128 (1991).
\bibitem{Mustafa} M.~G.~Mustafa,
Phys. Rev. C \textbf{72}, 014905 (2005).
\bibitem{collBitno2} M. G. Mustafa and M. H. Thoma, Acta Phys. Hung. A {\bf 22}, 93 (2005).
\bibitem{DjG} M. Djordjevic and M. Gyulassy, Nucl. Phys. A {\bf 733}, 265 (2004).
\bibitem{DCE3} B.~W.~Zhang, E.~Wang and X.~N.~Wang,
Phys. Rev. Lett. \textbf{93}, 072301 (2004).
\bibitem{MradDB} M.~Djordjevic and U.~Heinz,
Phys. Rev. C \textbf{77}, 024905 (2008).
\bibitem{rad0} M.~Djordjevic,
Phys. Lett. B \textbf{763}, 439-444 (2016).
\bibitem{DREENAC} D.~Zigic, I.~Salom, J.~Auvinen, M.~Djordjevic and M.~Djordjevic,
J. Phys. G \textbf{46}, no.8, 085101 (2019).
\bibitem{DREENAB} D.~Zigic, I.~Salom, J.~Auvinen, M.~Djordjevic and M.~Djordjevic,
Phys. Lett. B \textbf{791}, 236-241 (2019).
\bibitem{BjorkenH} J. D. Bjorken, Phys. Rev. D {\bf 27}, 140 (1983).
\bibitem{DREENAA} D.~Zigic, I.~Salom, J.~Auvinen, P.~Huovinen and M.~Djordjevic,
[arXiv:2110.01544 [nucl-th]].
\bibitem{ID} Z. B. Kang, I. Vitev and H. Xing, Phys. Lett. B {\bf 718}, 482 (2012); R. Sharma, I. Vitev and B. W. Zhang, Phys. Rev. C {\bf 80}, 054902 (2009).
\bibitem{Rad} M. Djordjevic, Phys. Rev. C \textbf{80}, 064909 (2009);
M. Djordjevic and U. Heinz, Phys. Rev. Lett.  \textbf{101},
022302 (2008).
\bibitem{DELF} M.~Djordjevic and M.~Djordjevic,
Phys. Lett. B \textbf{734}, 286-289 (2014).
\bibitem{MGF} M. Gyulassy, P. Levai and I. Vitev, Phys. Lett. B {\bf 538}, 282 (2002).
\bibitem{Dainese} A. Dainese, Eur. Phys. J. C {\bf 33}, 495 (2004).
\bibitem{nonC} M. Djordjevic, M. Djordjevic and B. Blagojevic, Phys. Lett. B {\bf 737}, 298 (2014).
\bibitem{HeavyBare} M.~Djordjevic, M.~Gyulassy, R.~Vogt and S.~Wicks,
Phys. Lett. B \textbf{632}, 81-86 (2006).
\bibitem{HeavyBareC} M.~Djordjevic,
Phys. Rev. Lett. \textbf{112}, no.4, 042302 (2014).
\bibitem{HeavyBareC1} M.~Djordjevic and M.~Djordjevic,
Phys. Rev. C \textbf{90}, no.3, 034910 (2014).
\bibitem{Kapusta}  J. I. Kapusta, {\it Finite-Temperature Field Theory}, Cambridge University Press  (1989).
\bibitem{Magm} M. Djordjevic, Phys. Lett. B {\bf 709}, 229 (2012).
\bibitem{bsga} B. Blagojevic, M. Djordjevic and M. Djordjevic, Phys. Rev. C {\bf 99}, no. 2, 024901 (2019).
\bibitem{HFP1} M.~Djordjevic,
Phys. Rev. C \textbf{85}, 034904 (2012).
\bibitem{masstomography} M.~Djordjevic, B.~Blagojevic and L.~Zivkovic,
Phys. Rev. C \textbf{94}, no.4, 044908 (2016).
\bibitem{51} M.~Djordjevic and M.~Djordjevic,
Phys. Rev. C \textbf{92}, no.2, 024918 (2015).
\bibitem{confirm1} V. Khachatryan {\it et al.} [CMS Collaboration], JHEP {\bf 1704}, 039 (2017); S. Acharya {\it et al.} [ALICE Collaboration], JHEP {\bf 1811}, 013 (2018).
\bibitem{xb1} A. Nakamura, T. Saito and S. Sakai, Phys. Rev. D {\bf 69}, 014506 (2004).
\bibitem{xb2} Yu. Maezawa {\it et al.} [WHOT-QCD Collaboration], Phys. Rev. D {\bf 81} 091501 (2010).
\bibitem{Bleicher} T.~Lang, H.~van Hees, J.~Steinheimer, G.~Inghirami and M.~Bleicher,
Phys. Rev. C \textbf{93}, no.1, 014901 (2016).
\bibitem{PollMarlene} P.~B.~Gossiaux, J.~Aichelin, T.~Gousset, M.~Nahrgang, V.~Ozvenchuk and K.~Werner,
Nucl. Phys. A \textbf{931}, 581-585 (2014).
\bibitem{KcollH} R.~S.~Kolevatov,
[arXiv:0812.0691 [hep-ph]]; R.~Kolevatov and U.~A.~Wiedemann,
[arXiv:0812.0270 [hep-ph]].
\bibitem{collAyala} A.~Ayala, J.~Magnin, L.~M.~Montano and E.~Rojas,
Phys. Rev. C \textbf{77}, 044904 (2008).
\bibitem{PathD} M. Djordjevic, D. Zigic, M. Djordjevic and J. Auvinen, Phys. Rev. C {\bf 99}, no.6, 061902 (2019).
\bibitem{RAAobs1} T. Renk, Phys. Rev. C {\bf 85}, 044903 (2012).
\bibitem{RAAobs2} D. Molnar and D. Sun, Nucl. Phys. A {\bf 932}, 140 (2014); {\bf 910-911}, 486 (2013).
\bibitem{BDMS} R.~Baier, Y.~L.~Dokshitzer, A.~H.~Mueller and D.~Schiff,
JHEP \textbf{09}, 033  (2001).
\bibitem{TG} M. H. Thoma and M. Gyulassy, Nucl. Phys. B {\bf 351} (3), 491-506 (1991).
\bibitem{CMS_jpsi} A.~M.~Sirunyan \textit{et al.} [CMS],
Eur. Phys. J. C \textbf{78}, no.6, 509 (2018).
\bibitem{ALICE_D} S.~Acharya \textit{et al.} [ALICE],
JHEP \textbf{10}, 174 (2018).
\bibitem{Lop} Abramowitz, M. and Stegun, I. A. (Eds.). {\it Handbook of Mathematical Functions with Formulas, Graphs, and Mathematical Tables}, 9th printing. New York: Dover, p. 13 (1972).
\end{references}
\end{document}